\documentclass[twocolumn,prb,aps,superscriptaddress, amsmath, amssymb]{revtex4}
\usepackage{color}
\usepackage{graphicx}
\usepackage{natbib}
\usepackage{subfigure}
\newcommand{\nc}{\newcommand}
\nc{\on}{\operatorname}
\nc{\wt}{\widetilde}
\nc{\Wick}{{\mathbb :}}
\nc{\R}{{\mathbb R}}

\newcommand{\beq}{\begin{equation}}
\newcommand{\eeq}{\end{equation}}
\newcommand{\bmul}{\begin{multline}}
\newcommand{\emul}{{\end{multline}}}
\newcommand\beqa{\begin{eqnarray}}
\newcommand\eeqa{\end{eqnarray}}
\newcommand\bea{\begin{array}}
\newcommand\eea{\end{array}}
\newcommand\ba{\begin{array}}
\newcommand\ea{\end{array}}

\newcommand{\neqa}{\nonumber\end{eqnarray}}

\newcommand{\D}{{\mathcal D}}

\nc{\CH}{{\mathcal H}}
\nc{\Db}{{\bar D}}
\nc\comment[1]{}

\nc{\CM}{{\mathcal M}}
\nc{\CN}{{\mathcal N}}

\newcommand{\re}{\relax{\rm I\kern-.18em R}}

\nc{\meV}{{\mathrm{\,meV}}}
\nc{\cG}{{\mathcal G}}

\renewcommand{\)}{\right)}

\comment{ \)}
\renewcommand{\bar}{\overline}

\nc{\al}{{\alpha}}

\begin{document}

\title{Theory of surface-induced multiferroicity in magnetic materials, thin films and multilayers}
\author{Ahmed R. Tarkhany}
\affiliation{Department of Physics and Centre for the Science of Materials, Loughborough University, Loughborough LE11 3TU, UK.}
\author{Marco Discacciati}
\affiliation{Department of Mathematical Sciences, Loughborough University, Loughborough LE11 3TU, UK.}
\author{Joseph J. Betouras}
\email{J.Betouras@lboro.ac.uk} 
\affiliation{Department of Physics and Centre for the Science of Materials, Loughborough University, Loughborough LE11 3TU, UK.}

\begin{abstract}
We present a theoretical study of the onset of electric polarization close to a surface in magnetic materials and in thin films and multilayers. We consider two different paths that lead to the onset of multiferroic behavior at the boundary in materials that are bulk collinear ferromagnets or antiferromagnets. These two paths are distinguished by the presence or absence of a surface induced Dzyaloshinskii-Moriya interaction which can be taken into account through Lifshitz invariants in the free energy of the system. Experimental consequences are discussed in the light of the developed theory.

\end{abstract}
\maketitle

\section{Introduction.} 

Magnetoelectric multiferroic materials  continue to attract much interest, due to both scientific as well as technological importance \cite{Khomskii, Wang, lottermoser}. 
Typically
the ferroelectric transition temperature is much higher than the magnetic one and a coupling between the two order parameters is weak. Representative examples are the
transition metal perovskites BiFeO$_3$, BiMnO$_3$, belonging to type I class of multiferroics (more details on the classification \cite{Khomskii, cheong, eerenstein, Ramesh, ederer}). 
When the two ordering temperatures are close or even coincide, such as in TbMnO$_3$ \cite{kimura} or TbMn$_2$O$_5$ \cite{hur} strong multiferroic behavior is expected.
From the symmetry point of view, the necessity to break both inversion and time-reversal symmetries suggests different possible mechanisms which have been actually realised. In type II multiferroics, magnetism drives the onset of the ferroelectric order parameter, either due to the presence of spin-orbit coupling (SOC) and magnetic frustration 
e.g. in  Ni$_3$V$_2$O$_6$  \cite{lawes, mostovoy, katsura}, or exchange striction 
e.g. in TbMnO$_3$ and Ca$_3$CoMnO$_6$  \cite{sergienko, Wu} or "phase dislocated" spin density waves e.g. in YMn$_2$O$_5$ \cite{chapon, betouras}.

Currently the role of surfaces and interfaces in the properties of materials is the focus of systematic studies \cite{Panagopoulos1, Gerhard, Maruyama, Zutic, Hu}.  In technological applications there are important prospects and the theoretical understanding is developing \cite{Hellman, Sander}.
Experimental techniques have been advanced such that novel phenomena can be detected as a result of the higher precision and resolution.  Recent advancements led to the detection of new properties, by distinguishing surface from bulk phenomena or going to the atomic scale \cite{UO2, Sonntag}.

In this work, we study the effects of boundaries in the development of multiferroic behavior in bulk materials, thin films and multilayers. This is a complementary effort to first-principles calculations on the magnetoelectric coupling close to surfaces  \cite{Tsymbal} or monolayers \cite{Nakamura} of specific materials. A straightforward Ginzburg-Landau (GL) free-energy analysis with appropriate boundary conditions demonstrates that collinear magnetism can generate a ferroelectric polarization near surfaces, even without invoking the mechanism of phase dislocation \cite{betouras}. In addition, due to the absence of inversion symmetry close to surfaces, a term that promotes the Dzyaloshinskii-Moriya interaction (DMI) can be present, leading to multiferroic behaviour through the formation of spiral magnetic order \cite{mostovoy, katsura}. The underlying assumption is that we deal with predominantly magnetic materials with non-zero coupling between magnetic and ferroelectric order parameters. In the following, we analyse separately the two mechanisms.

\section{Ginzburg-Landau analysis.} 

We focus on a simple cubic ferromagnet, for simplicity, without frustration. Each spin has an interaction with its six nearest neighbours, according to an isotropic ferromagnetic Heisenberg-type interaction $J$ and the expectation value of the z-component of the magnetisation (spin) at site ${\bf l}$, $ m({\bf l})=\langle S_z({\bf l}) \rangle /S$ is the order parameter of the system.  The crystal is assumed to have a (001) surface.
In the continuous space approximation, the GL equation of the magnetization $m({\bf l})$ which depends only on $z$, reads \cite{Mills} (for completeness the details are in Appendix A):

\begin{equation} 
\frac{a_0^2}{6}\frac{\partial^2m(z)}{\partial z^2} + (1 - \tau)m(z) - \beta m^3(z)=0
\end{equation}
\noindent where $a_0$ is the lattice parameter and  $\tau =\frac{T}{T_c}$ is the reduced temperature ($T_c$ is the Curie temperature) and $\beta = \frac{3}{5}[s(s+1) + \frac{1}{2}]/(s+1)^2$. In the limit of $(z\to\infty)$, far from the surface, the order parameter takes the bulk value $m_{\infty} = (1 - \tau)^{1/2}/\beta^{1/2}$. The free energy expansion, leads to the same equation for both a ferromagnet or an antiferromagnet \cite{Mills}.
Defining $m(z) \equiv m_{\infty}f(z)$ and $\xi^2 \equiv \frac{1}{6}\frac{a_0^2 }{(1 - \tau)}$ then:

\beq  \xi^2\frac{\partial^2f(z)}{\partial z^2} + f(z) - f^3(z) = 0 \eeq

The boundary conditions are $f(\infty)=1$ and $f(0)=a_0 \frac{\partial f}{\partial z} (0)$. Then the solution of Eq. (1) reads:
$m(z) = m_{\infty} \tanh (\frac{z+a_0}{\sqrt{2}\xi})$.

\noindent 
We now address the question whether the onset of ferroelectricity is possible due to the existence of this surface which results in the change of the magnetisation. The physical argument is that as the inversion symmetry is broken due to the surface, the onset of multiferroic behavior is possible.

The free energy is supplemented by a term due to the coupling between magnetization and electric polarization {\bf p} and a term which is the electronic part of the free energy that depends only on {\bf p}. Since we are interested in systems where the magnetic order is the primary one, the second term is sufficient to be quadratic in {\bf p}.  The free energy reads:

\begin{eqnarray}
\nonumber
\delta F = F_{ME} + F_{E} = {\bf p} \cdot (\gamma \nabla ({\bf m}^2) \\
+ \gamma \prime [{\bf m}(\nabla \cdot {\bf m}) - ({\bf m} \cdot \nabla) {\bf m}] + ...) + \frac{{\bf p}({\bf r})^2}{2 \chi_E}
\end{eqnarray}
 
Note that, in the case of collinear magnetic structure the term proportional to $\gamma$ is non-zero, while the term proportional to $\gamma'$ (a Lifshitz invariant for cubic lattice) is zero. Taking the dielectric susceptibility as constant and using $m(z)$ as the magnetic order parameter, then the minimisation with respect to $p(z)$ results in:
 \begin{eqnarray}
 \nonumber
 p(z) = - \chi_E(\gamma \nabla ({\bf m}^2) + \gamma \prime \left[{\bf m}(\nabla \cdot {\bf m}) - ({\bf m} \cdot \nabla) {\bf m}\right] + ...)
 \end{eqnarray}
 
 \noindent Using the solution of Eq. (1) for the magnetisation, $p(z)$ becomes:
 
 \beq p(z) = - m_{\infty}^2 \frac{\sqrt{2}}{\xi} \chi_E\gamma\left[\tanh(\frac{z+a_0}{\sqrt{2}\xi}) [1- \tanh^2(\frac{z+a_0}{\sqrt{2}\xi})] \right] \eeq
  
 \begin{figure}[t!]
   \begin{center}
    {\label{}\includegraphics[scale=0.34]{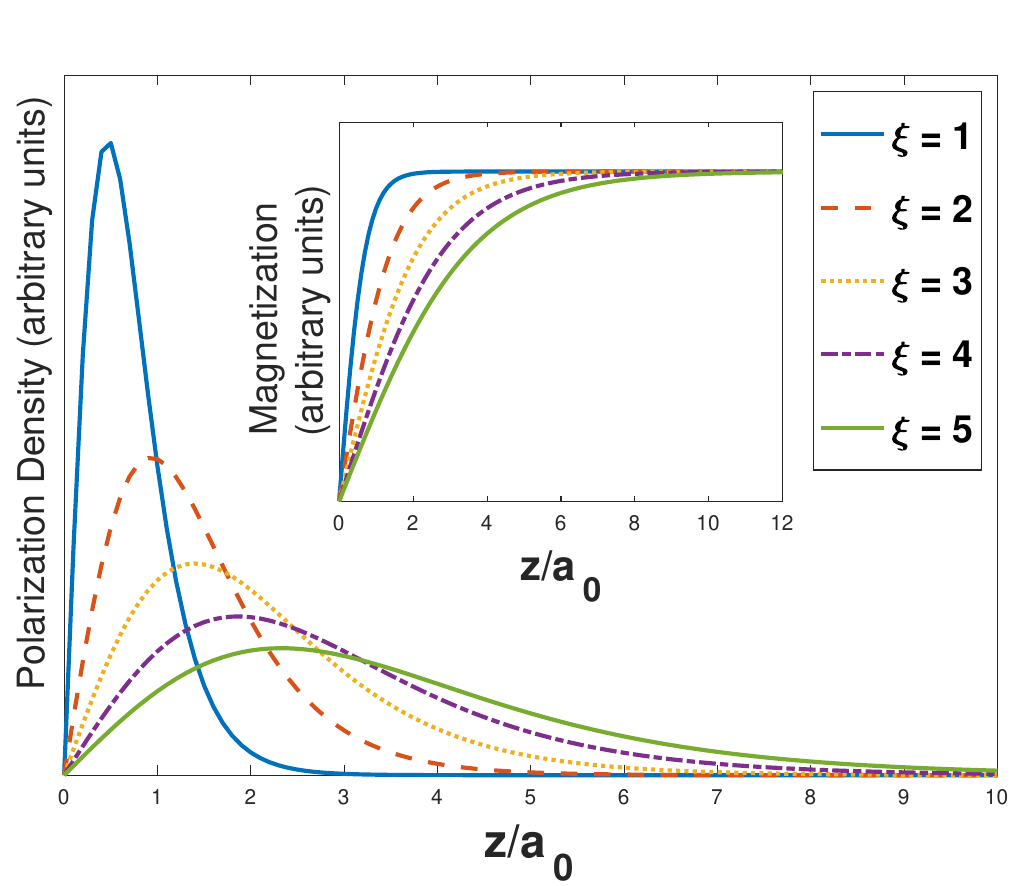}} \\
  \end{center}
  \caption{Polarization as a function of the distance from the surface.}
  \label{fig:polarization_1}
\end{figure}

This function is plotted in Fig. 1, interestingly it is peaked at a distance from the surface $z=\sqrt{2}\xi \tanh ^{-1}(1/\sqrt{3})$ with  $p_{max} = -  \frac{2 \sqrt{2}}{3 \sqrt{3}} \frac{1}{\xi}\chi_E \gamma m_{\infty}^2 \approx - \frac{1}{\xi}\chi_E \gamma m_{\infty}^2 $, where the negative sign denotes a direction opposite to the direction of $\nabla ({\bf m}^2)$.
 The distance over which polarization is developed and the location of its peak is also controlled by the temperature through the magnetic correlation length. 
For a reasonable estimate of the effect we need the range of possible values of $\gamma \chi_E$ which in the literature \cite{Coupling1,Coupling2,Coupling3,Coupling4} can be found to lie between 10$^{-23}$ and 10$^{-13}$ sm/A, a typical correlation length \cite{Michels} $\xi \approx$ 10 nm and a typical value for the bulk magnetization $m_{\infty} \approx$ 10-100 kA/m. Then the range of the possible values of  $p_{max}$ is between 10$^{-10}$ to 10$^2$ $\mu$C/cm$^2$ or to produce a measurable polarisation a value for $\gamma \chi_E$ larger than 10$^{-19}$ sm/A is enough \cite{explanation}.

 
 %
 %
 
 \section{DMI in magnetic thin films and multilayers.}

The lack of inversion symmetry close to surfaces can lead to the induction of DMI.
The direction of the {\bf d} vector in this case can cause the directions of the spins of the nearest neighbours to change in such a way as to break the chiral symmetry close to the surface  \cite{xia, crepieux}.
For the purpose of our investigation of the chiral nature induced by the DMI, the antisymmetric exchange interaction is described by a Lifshitz invariant
term which is linear in the spatial derivatives of the magnetisation {\bf m(r)} of the form $m_i \frac{\partial m_j}{\partial x_k} - m_j \frac{\partial m_i}{\partial x_k}$, where $x_l$ denotes a spatial coordinate.
These interactions are responsible for breaking the chiral symmetry and stabilizing
localized magnetic vortices, with certain chirality which has been observed experimentally in
noncentrosymmetric ferromagnetic and antiferromagnetic materials \cite{examples}. It is possible to
observe these effects in centrosymmetric crystals where stresses or applied magnetic fields \cite{Bogdanov-2001} or anisotropic frustrated magnetic interactions \cite{leonov}  induce chiral magnetic couplings and vortices/skyrmions. Chiral effects, as a consequence of 
the DMI energy are not so strong in the bulk, but they can become fundamentally important in magnetic thin films and multilayers or near
the surface of a larger crystal where the local symmetry is low.
Taking into account experimetal facts, the chiral couplings should also be inhomogeneous  \cite{Bogdanov-2001}
within a magnetic structure with low local symmetry. A phenomenological term for the corresponding chiral energy density is ${\mathcal F}_D = D f({\bf r}) {\mathcal L}({\bf m})$
where $D$ is a constant, ${\mathcal L}$ is a Lifshitz invariant and the
function $f({\bf r})$ is a function describing the inhomogeneous distribution of the
magnetic chiral energy. $f({\bf r})$ was interpreted as another field,
in addition to the magnetisation \cite{Bogdanov-2001}, but it essentially indicates the strength profile of the DMI as a function of the distance from the surface.  To demonstrate the physics clearly, we take two functions as the profile function $f(z)$, where $z$ is the distance from the surface: a function exponentially decaying in $z$ and a function $1-tanh(z/\lambda)$, both with maximum at the surface. This behavior has been verified when the magnetisation in a finite-width slab was computed \cite{Bogdanov-2001}. 

We consider a uniaxial magnetic anisotropy that contributes to the energy density a term ${\mathcal F}_{an} = - K m_z^2$.  Then the free energy density reads:
\begin{eqnarray}
{\mathcal F} = A \sum_i \left(\frac{\partial{\bf m}}{\partial x_i}\right)^2 + D f(r) {\mathcal L}  - K m_z^2 +{\mathcal F}_{ME} + {\mathcal F}_{E}
\end{eqnarray}

\noindent the first term represents the magnetic exchange interaction with a stiffness constant A (in , the second is 
the energy of the electric polarization with susceptibility $\chi_E$, the third term is due to DMI coupling, the fourth due to anisotropy and ${\mathcal F}_{ME}$ is the term of the free energy that couples ${\bf m}$ and ${\bf p}$. In the following all the quantities are dimensionless and for that purpose the magnetization $\vec{m}$ is normalized by its amplitude $M_0$, D is in units of stiffness A per m, $chi_E$ is in units of vacuum permittivity $\epsilon_0$, K is in units of A per m$^2$, polarization p in units of $\sqrt{\epsilon_0 \mu_0} M_0$ where $\mu_0$ is vacuum permeability, length is in units of lattice constant, $\gamma$ is in units of inverse polarization and the coupling constant in ${\mathcal F}_{ME}$ is the product D $\gamma$ (instead of only $\gamma$).

\begin{figure}[t!]
\subfigure[]{\includegraphics[width=0.48\columnwidth, trim= 0 0 0 0, clip]{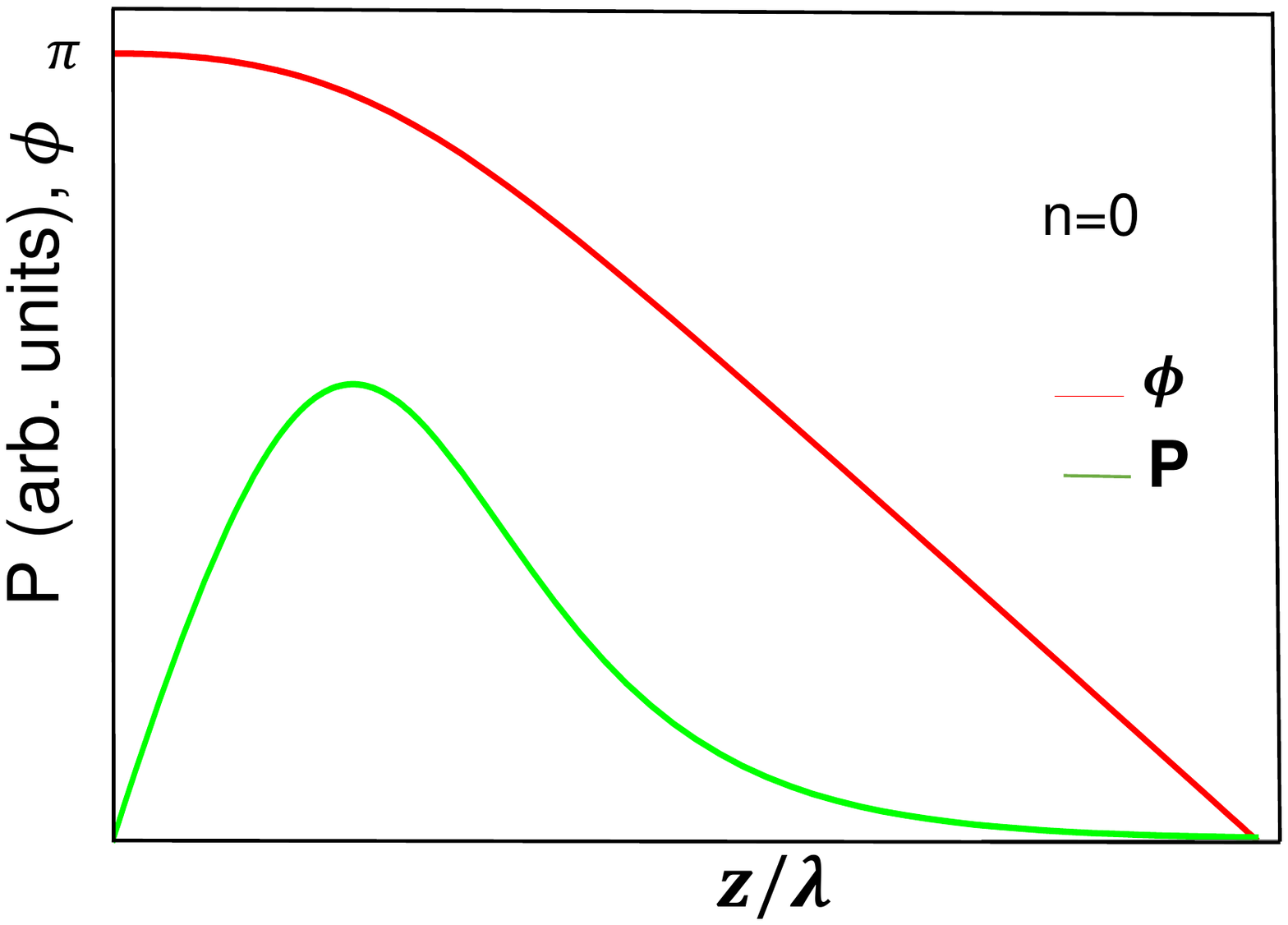}\label{fig:fig1label}}
\subfigure[]{\includegraphics[width=0.49\columnwidth, trim= 0 0 0 0, clip]{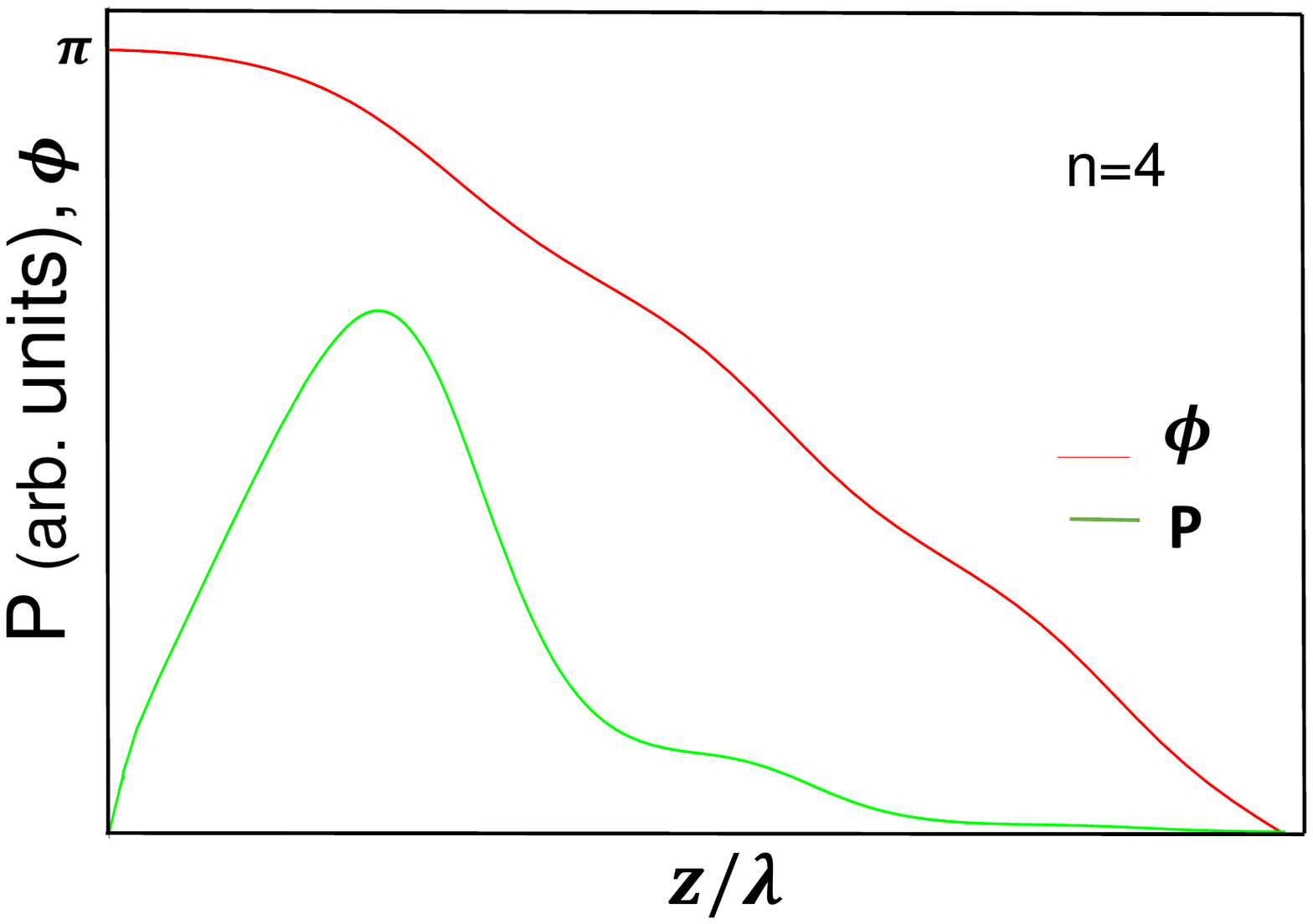}\label{fig:fig2label}}
\caption{magnetisation ($\phi$) and polarization as a function of the distance from the surface for in-plane anisotropy with $n=0$ (a) and $n=4$ (b). The used parameters are A=1, D=1, k=1, $\gamma$=1, $\chi_E$=1.5. The peak of the polarization is at a distance from the surface in agreement with Fig. 1.}
\label{fig:polarization_2}
\end{figure}

\subsection{Anisotropic term with $K<0$}

When $K < 0$ in Eq. (5) then $\bf{m}$ lies in the $xy$ plane. The Lifshitz invariant ${\mathcal L}$ can be taken as $(m_x \frac{d m_y}{dz} - m_y \frac{d m_x}{dz})$. The free energy term that couples ${\bf m}$ and ${\bf p}$ then takes the form
${\bf p} \cdot ( {\bf m} \times  \frac{d {\bf m}}{dz})$ which in this particular geometry reads: $D \gamma p_z (m_x \frac{d m_y}{dz} - m_y \frac{d m_x}{dz})$.
It is convenient to work with the angle $\phi$ to describe the vector ${\bf m}$ that lies in the $xy$ plane. The related part of the free energy density becomes:
\begin{eqnarray}
\nonumber
 {\mathcal F} = A(\frac{d\phi}{dz})^2 + D f(z) (1 + \gamma p_z) \frac{d\phi}{dz} +\frac{p_z^2}{2 \chi_E} - k cos(n\phi).
 \end{eqnarray}
\noindent The last term describes an in-plane anisotropy with $n$ an even integer, depending on lattice symmetry and/or
homogeneous strain \cite{Bogdanov-2001}. Minimizing the free energy with respect to ${\bf m}$ and ${\bf p}$, we obtain:
\begin{eqnarray}
\nonumber
 2A\frac{d^2\phi}{dz^2} + D \frac{df}{dz} (1 + \gamma p_z) &&+ D f \gamma \frac{dp_z}{dz} -nksin(n\phi) \\
 &&= 0 \\
 p_z&&= - D \gamma \chi_E f \frac{d\phi}{dz} 
\end{eqnarray}


Inserting Eq. (7) into Eq. (6), we solve numerically for $\phi$  and $p_z$ as a function of the distance from the surface. The results are presented in Fig. 2. This physics provides a second mechanism to generate a finite polarization close to the surface which comes from the non-zero value of $\frac{d\phi}{dz}$ as a consequence of the DMI which is maximum at the surface. $f(z)$ is the profile that controls the strength of the DMI. We have checked that both profiles of $f(z)$ lead to the same physics qualitatively and we have used $1-tanh(z/\lambda)$ in Figs 2 and 3.


\begin{widetext}

\begin{figure}[htp]
\subfigure[]{\includegraphics[height=1.4in, width=0.4\columnwidth, trim= 0 0 0 0, clip]{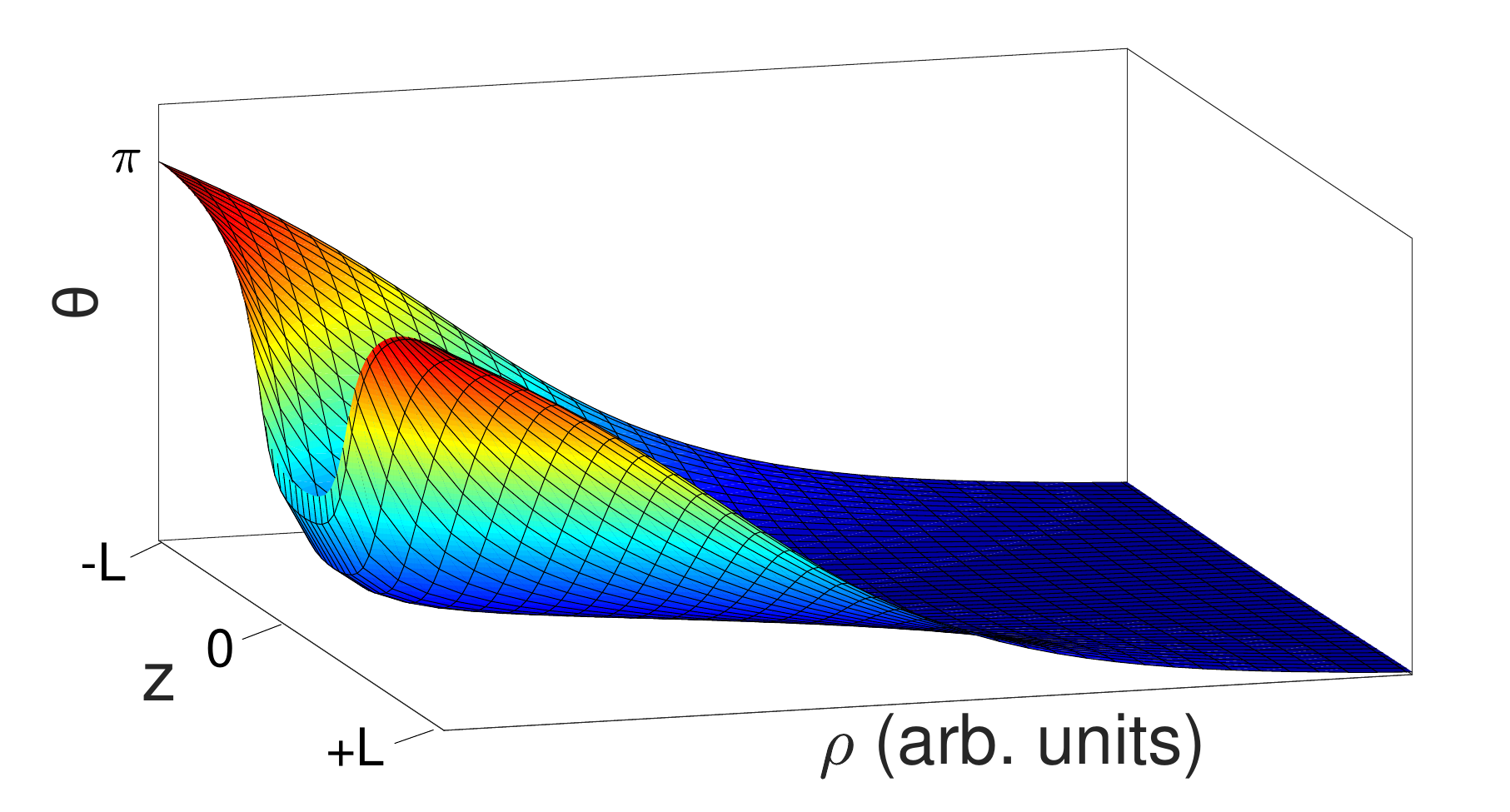}\label{fig:fig1label}}
\subfigure[]{\includegraphics[height=1.4in, width=0.55\columnwidth, trim= 0 0 0 0, clip]{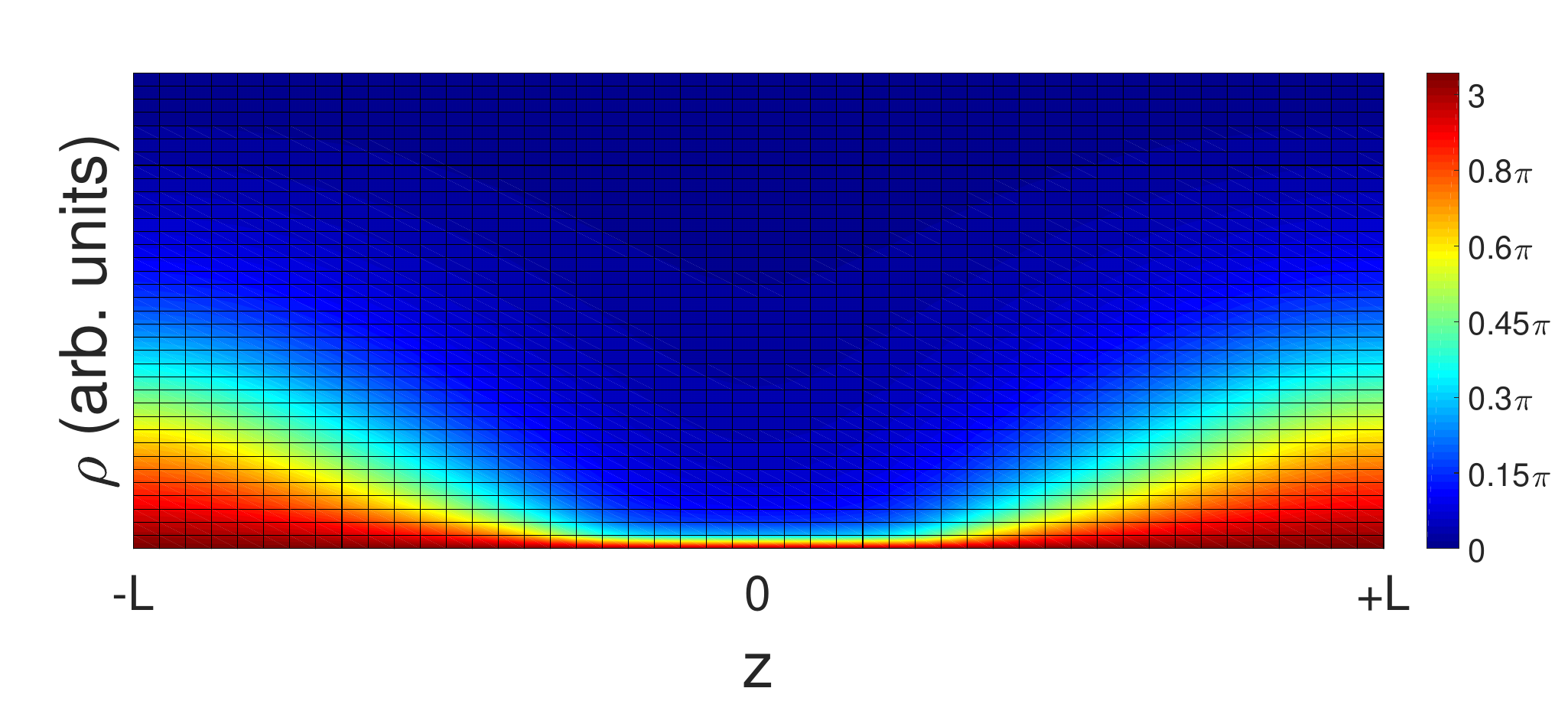}\label{fig:fig2label}}
\subfigure[]{\includegraphics[height=1.3in, width=0.38\columnwidth, trim= 0 0 0 0, clip]{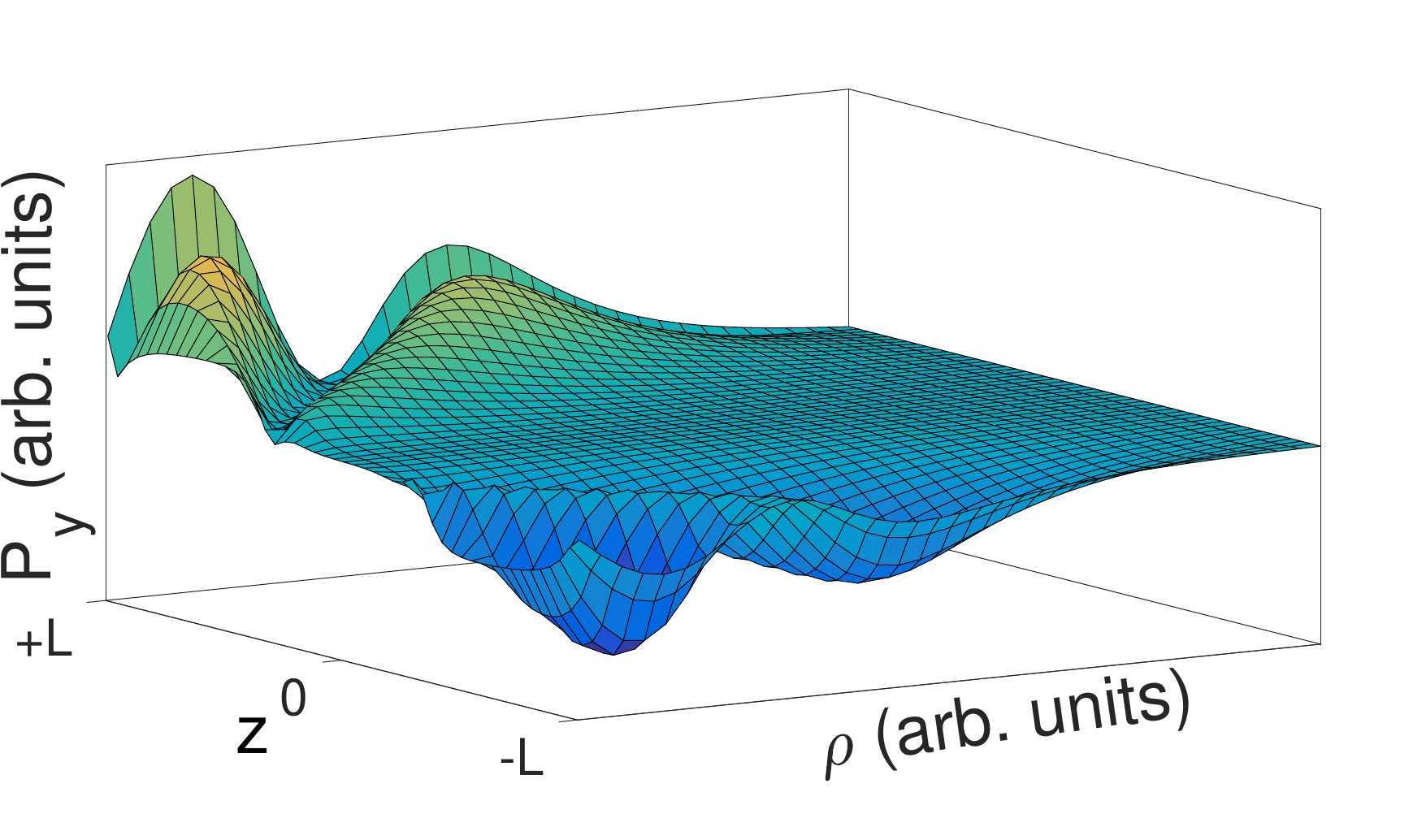}\label{fig:fig3label}}
\subfigure[]{\includegraphics[height=1.25in, width=0.23\columnwidth, trim= 0 0 0 0, clip]{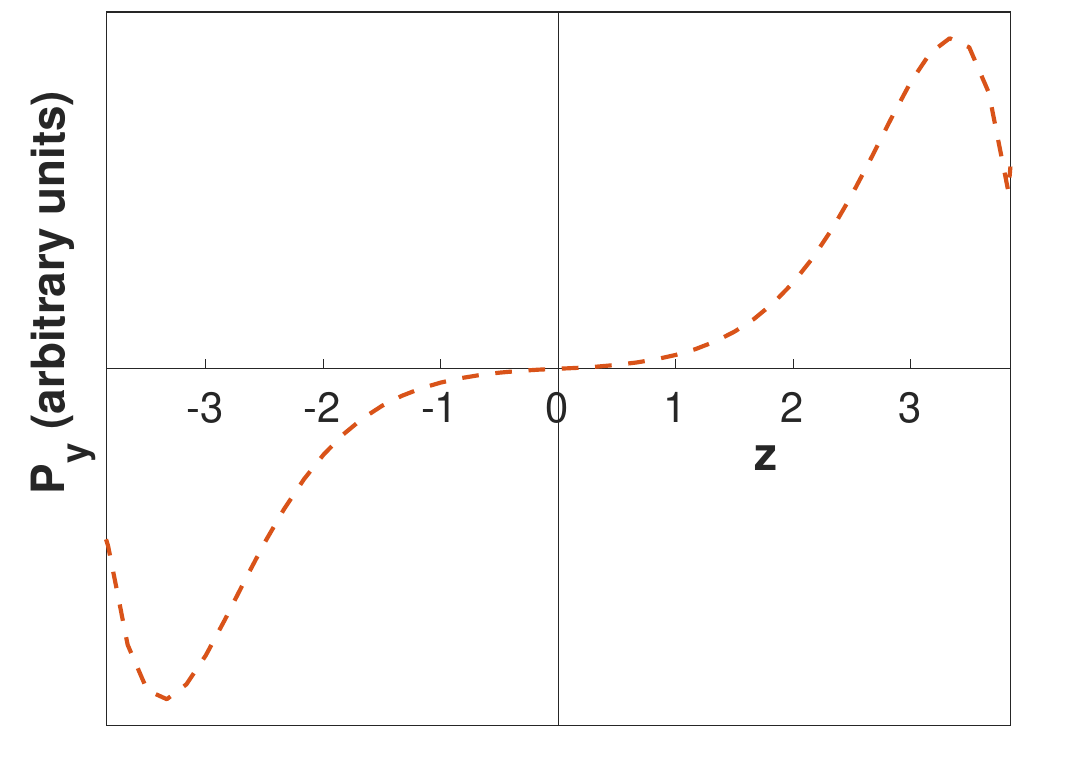}\label{fig:fig4label}}
\subfigure[]{\includegraphics[width=0.37\columnwidth, trim= 0 20 0 0, clip]{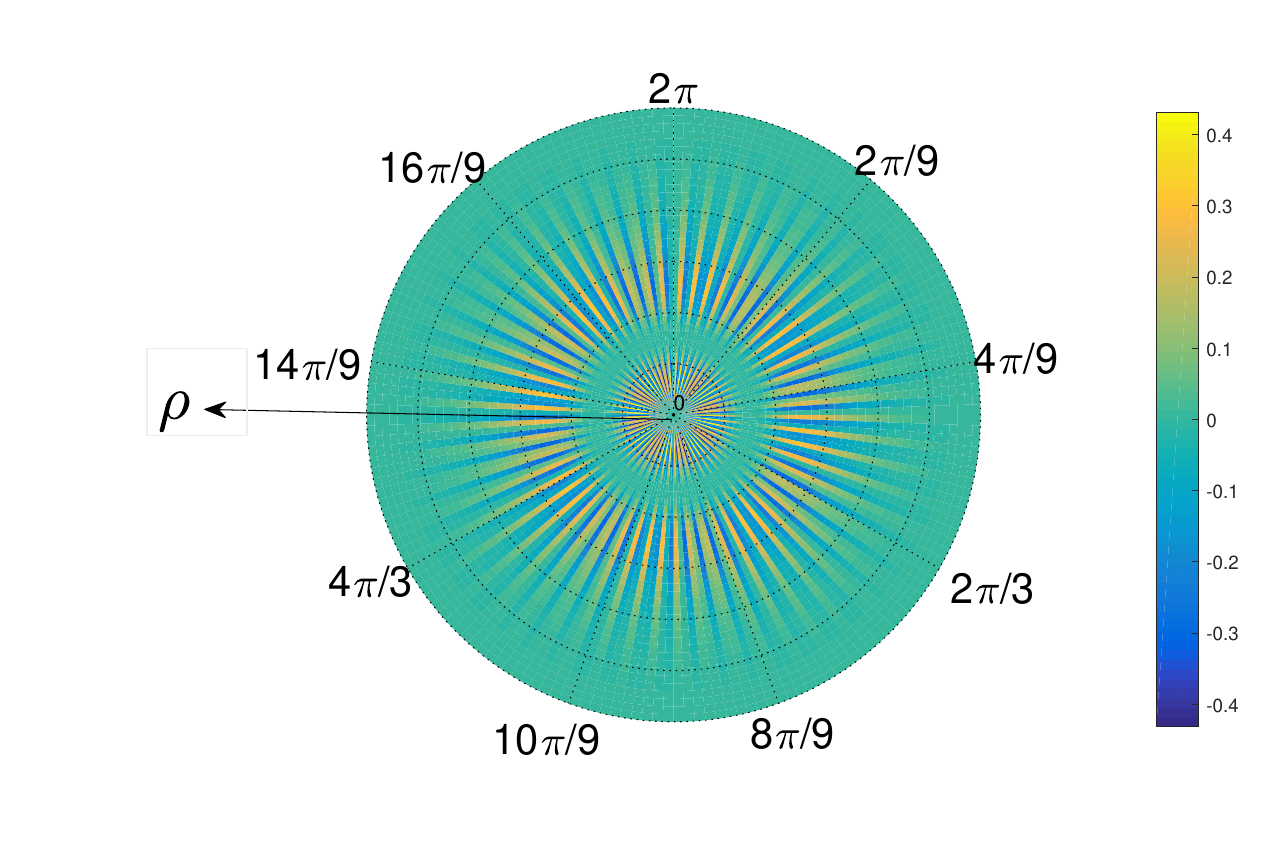}\label{fig:fig1label}}
\caption{Numerical results for $K=1$. In (a) and (b) $\theta$ as a function of $z$ and $\rho$ with boundary conditions as explained in the text. The surface are at z=L and z=-L, while the bulk is at z=0. (c) The component of polarization $p_y$ as a function of $z$ and $\rho$, (d) $p_y$ at fixed $\rho$=  0.05 R and (e) at fixed z= 0.96 L (and L=4 $\lambda$), where it shows oscillatory behavior as a function of $\rho$ and $\phi$ for the second layer from the surface. The x-component of the polarization is out of phase by $3\pi/2$ in $\phi$. The rest of the parameters are the same as in Fig. 2.}
\label{fig:Kpositive}
\end{figure}

\end{widetext}

For an estimate of the effect, the range of values of the product $D \gamma \chi_E$ is taken between 10$^{-23}$ and 10$^{-13}$ sm/A, 
the  intralayer spacing $\Delta z \approx  \cal{O}$ (1 nm),  typical values for the magnetization \cite{Panagopoulos1, Boulle, Takeuchi} $M_0 \approx$ (10-1000) kA/m and typical $\Delta \phi \approx 0.1$ so that $\frac{\Delta \phi}{\Delta z} \approx$ 10$^5$ m$^{-1}$ while $f$ is of $\cal{O}$(1) close to the surface. Then the range of the possible values of the maximum polarization is $p_z \approx$ 10$^{-10}$ to 10$^4$ $\mu$C/cm$^2$. Again with a reasonable value for $\D\gamma \chi_E$ of 10$^{-17}$ sm/A or larger, the polarization can be detected \cite{explanation}.

\subsection{ Anisotropic term with $K>0$}

When $K>0$ in the free energy, the relevant Lifshitz invariants may involve gradients along all three directions depending on the respected symmetry. There is a  Lifshitz invariant term ${\mathcal L}$ purely magnetic as well as a term ${\mathcal L}_{me}$ that mixes ${\bf m}$ and ${\bf p}$ in the free energy.  In the case of twofold or fourfold symmetry about the z axis we take the Lifshitz invariants to be: 

\begin{widetext}
\begin{equation}
{\mathcal L}+{\mathcal L}_{me} =  m_z \frac{\partial m_x}{\partial x} -  m_x \frac{\partial m_z}{\partial x} + m_z \frac{\partial m_y}{\partial y}  - m_y \frac{\partial m_z}{\partial y}  
+   \gamma m_z (p_x  \frac{\partial m_y}{\partial z} - p_y \frac{\partial m_x}{\partial z}) 
 \end{equation}
\end{widetext}

%

Using spherical coordinates for the magnetisation ${\bf m}= (sin\theta cos\phi, sin\theta sin\phi, cos\theta)$, cylindrical coordinates for the spatial vector ${\bf r}=(\rho cos\zeta,\rho sin\zeta,z)$ and focusing on the magnetic part of the free energy, the problem has axisymmetric localized solutions $\phi=\zeta$ and $\theta = \theta (\rho,z)$ with $\theta(0)= \pi$ and $\theta (\infty)=0$. The part of the free energy proportional to  ${\mathcal L}_{me}$ reads:
%
${\mathcal F}_{me}= D f(z) \gamma \cos^2\theta \frac{\partial\theta}{\partial z} \left(p_x \sin\phi - p_y  \cos\phi \right)$.   
%
Minimizing the free energy with respect to $\theta$ and $p_x$, $p_y$, we obtain:

\begin{widetext}
\begin{eqnarray}
A\left[\frac{\partial^2\theta}{\partial z^2}+\frac{\partial^2\theta}{\partial \rho^2}+\frac{1}{\rho}\frac{\partial \theta}{\partial \rho} - \frac{sin\theta cos\theta}{\rho^2}\right] -D f(z) 
\frac{sin^2\theta}{\rho} -Ksin\theta cos\theta =0 \\
p_{x} = - \chi_E D f \gamma  \sin\phi \cos^2\theta \frac{\partial\theta}{\partial z}\;\;\; and \;\;\;\;  p_{y} =  \chi_E D f \gamma  \cos\phi \cos^2\theta  \frac{\partial\theta}{\partial z} 
\end{eqnarray}
\end{widetext}



\noindent As ${\bf p}$ is the sub-dominant order parameter, to simplify the calculation it is sufficient to neglect its effect on $\theta$.  Then the solution for ${\bf m}$ is similar to the purely magnetic problem \cite{Bogdanov-2001}. As a result, the magnitude of ${\bf p}$ at any given point is $p= \chi_E D f \gamma \cos^2\theta  |\frac{\partial\theta}{\partial z}|$. In Fig. (\ref{fig:Kpositive}) we present $\theta$ as a function of $\rho$ and $z$ as well as $p_x$ and $p_y$ (differing by a phase difference) (more details in Appendix B).  Similarly to the case (i), using similar typical values and  reasonable estimate of $|\frac{\partial \theta}{\partial z}| \approx$ 10$^4$ m$^{-1}$, then the range of the possible values of the amplitude of ${\bf p}$ is $p \approx$ 10$^{-8}$ to 10$^{4}$ $\mu$C/cm$^2$, which suggests a detectable value \cite{explanation}.

\section{Discussion.} 

We present a detailed study of the multiferroic behavior in thin films, multilayers and close to the surface of magnetic materials with symmetries  (inversion and time reversal) that do not allow the onset of polarization in the bulk. The onset of polarization can be achieved both in the presence or absence of a DMI close to the surface. One mechanism is through the non-zero gradient of the magnetization, while the second is through the change of orientation of the magnetization (and not its amplitude) as a function of the distance from the surface. Our study is essentially the inverse effect of Ref. (\onlinecite{valencia}), where surface-induced magnetisation was detected in the archetypal ferroelectric BaTiO$_3$. In the same spirit, there exist recent intensive efforts to synthesize multiferroic heterostructures and control the interfacial DMI (e.g. Refs. [\onlinecite{Vaz, Jain, hu, Ding, Hrabec, Kuepferling}]). In thin films  DMI originating from strong SOC of interfacial atoms neighboring the magnetic layer can be engineered and controlled. DMI has been engineered in a ferromagnet (FM) interfaced with two different heavy metals, such as Pt/Co/Ir \cite{Chen, Moreau-Luchaire}, or in magnetic layers inserted between a heavy metal and an oxide, such as Pt/CoFe/MgO \cite{Emori, Emori2}. In thin films, the deposition conditions would make the couplings vary, given that the strength of the DMI has been recently shown to vary in a model system Pt/Co/Pt, depending e.g. on the temperature variation during deposition \cite{Wells}. An elegant and reliable method for determining the magnitude of the DMI from static domain measurements even in the presence of hybrid chiral structures was recently demonstrated \cite{Legrand} while electrical detection of single magnetic skyrmions has been achieved at room temperature in metallic multilayers \cite{Maccariello}. 

We estimate the range of values of the polarization, that can be detected even if the coupling constant that mixes the magnetic and ferroelectric order parameters is several orders of magnitude smaller than the highest ones reported in the literature.  
The polarization we predict is detectable within the current accuracy of the experimental techniques. Techniques such as polarised neutron reflectivity can be used to detect magnetisation gradients over a length scale of a few nanometers. In case of out-of-plane components, magnetic force microscopy is ideal to detect the magnetisation components. X-ray magnetic circular/linear dichroism can detect the contribution of individual magnetic ions if required. 
Scanning transmission X-ray microscopy, has been used effectively to study skyrmion dynamics high temporal and spatial resolution \cite{Stoll}. Recently, the linear magnetoelectric phase in ultrathin MnPS$_3$ was probed by optical second harmonic generation \cite{Chu}.  These techniques in combination with e.g. electrostatic force microscopy or ellipsometry make feasible the detection of both order parameters.

\section*{Acknowledgments.} 

We thank  N. Banerjee, A. Bogdanov, P. Borisov, D. Efremov, N. Gidopoulos, T. Hesjedal, D. Khomskii, P. King, P. Radaelli,  I. Rousochatzakis and J. van den Brink for useful discussions and communications. The work is supported partly by EPSRC through the grant EP/P003052/1 (JJB) and a scholarship from the Regional Government of Kurdistan-Iraq (ART). JJB also thanks the Isaac Newton Institute for Mathematical Sciences for support and hospitality during the programme "Mathematical design of new materials", supported by EPSRC grant no EP/R014604/1, where a part of the work was done.

\section*{Appendix A: Derivation of the Ginzburg-Landau equation.}

Using molecular field theory, the expectation value of the z-component of the magnetisation (spin) at site ${\bf l}$, $ m({\bf l})=\langle S_z({\bf l}) \rangle /S$ which is the order parameter of the system, is written as:
\begin{equation}
m({\bf l})= B_s\left(- \frac{JS}{k_B T} \sum_{{\bf {\hat \delta}}} m({\bf l}+{\hat {\bf \delta}})\right)
\end{equation}
\noindent where $k_B$ is Boltzmann's constant, $T$ is the temperature, $S$ is the spin of the magnetic ions and $B_s(x)$ is the Brillouin function $B_s(x)=(1+\frac{1}{2S})\coth[(1+\frac{1}{2S})x] - \frac{1}{2S}\coth(\frac{x}{2S})$. The sum over ${\hat {\bf \delta}}$ ranges over the six nearest neighbours of the spin at site ${\bf l}$.
We can then expand $\coth(\theta)$ in powers of $\theta$: 
\begin{equation} 
 \coth{\theta} = \frac{1}{\theta} +\frac{\theta}{3} - \frac{\theta^3}{45} + .......
 \end{equation}
 The Brillouin function then reads: 
 \begin{equation}
 B_s(x) = \frac{(s+1)}{3s}\{x - \frac{1}{15s^2}[s(s+1) +\frac{1}{2}]x^3 +......\}.
 \end{equation}
We define the reduced temperature $\tau =\frac{T}{T_c}$ (for the model we discuss $T_c$ is the Curie temperature $T_c = 2k_BJ(S+1)/3S$).
The order parameter is then defined through the relations $ m({\bf l}) = <S({\bf l})>/S$ with $m({\bf l}) = B_s\left( - \frac{\sum{m({\bf l} +{\bf \delta})}}{2\tau k_B(s+1)}\right)$.
In the continuous space approximation of the lattice, $m({\bf l})$ becomes also function of the continuous ${\bf l}$ and 
\begin{equation}
\sum_{{\bf \delta}}{m({\bf l}+{\bf \delta})}  \simeq  6  m({\bf l})  + a_0^2 \nabla^2 m({\bf l})
\end{equation}
\noindent where $a_0$ is the lattice parameter.
Retaining only first order in $\bigtriangledown^2 m$ terms and using $\beta = \frac{3}{5}[s(s+1) + \frac{1}{2}]/(s+1)^2$ and the fact that  $m({\bf r})$ depends only on $z$, the GL equation becomes:
\begin{equation} 
\nonumber
\frac{a_0^2}{6}\frac{\partial^2m(z)}{\partial z^2} + (1 - \tau)m(z) - \beta m^3(z)=0
\end{equation}

\noindent which is Eq. (1) of the main text. The free energy expansion, leads to the same equation for both a ferromagnet or an antiferromagnet.

\begin{widetext}

\section*{Appendix B: Details of the calculation of magnetisation for K$>$0.}

As explained in the text, the magnetisation in the case K$>$0 is determined by the function $\theta=\theta(\rho,z)$ such that
\begin{equation}\label{eq:1}
  A \left( \frac{\partial^2 \theta}{\partial \rho^2} + \frac{\partial^2 \theta}{\partial z^2} + \frac{1}{\rho} \frac{\partial \theta}{\partial \rho} - \frac{\sin(\theta)\cos(\theta)}{\rho^2} \right) + D f(z) \frac{\sin^2(\theta)}{\rho} - K \sin(\theta)\cos(\theta) = 0 \quad \mbox{in } \Omega
\end{equation}
where $\Omega = (0,r) \times (-l,l)$. The boundary conditions are
\begin{itemize}
\item[-]
on $\Gamma_4 = \{ (\rho,z) \, : \, \rho = 0, \; -l \leq z \leq l \}$, we impose $\theta = \pi$
\item[-]
on $\Gamma_2 = \{ (\rho,z) \, : \, \rho = r, \; -l \leq z \leq l \}$, we impose $\theta = 0$
\item[-]
on $\Gamma_1 = \{ (\rho,z) \, : \, 0 \leq \rho \leq r, \; z = -l \}$ and on $\Gamma_3 = \{ (\rho,z) \, : \, 0 \leq \rho \leq r, \; z = l \}$, we impose periodic boundary conditions: $\theta(\rho,z=-l) = \theta(\rho,z=l)$.
\end{itemize}

Equation \eqref{eq:1} can be equivalently reformulated as
\begin{equation}\label{eq:2}
A \left( \frac{\partial^2 \theta}{\partial \rho^2} + \frac{\partial^2 \theta}{\partial z^2} + \frac{1}{\rho} \frac{\partial \theta}{\partial \rho} \right) - \frac{1}{2} \left( \frac{A}{\rho^2} + K \right) \sin(2\theta) + D\frac{f(z)}{\rho} \sin^2(\theta) = 0 \quad \mbox{in } \Omega \, .
\end{equation}
Defining the linear operator
\begin{equation}
L\theta = A \left( \frac{\partial^2 \theta}{\partial \rho^2} + \frac{\partial^2 \theta}{\partial z^2} + \frac{1}{\rho} \frac{\partial \theta}{\partial \rho} \right)
\end{equation}
and the nonlinear operator
\begin{equation}
f(\theta) = -\frac{1}{2} \left( \frac{A}{\rho^2} + K \right) \sin(2\theta) + D\frac{f(z)}{\rho} \sin^2(\theta)
\end{equation}
we can write the problem in abstract form as: find $\theta$ such that
\begin{equation}\label{eq:abstract}
L\theta + f(\theta) = 0 \quad \mbox{in } \Omega \, .
\end{equation}

\subsection*{Numerical solution using the finite difference method}

We introduce the computational grid as the one shown in the following figure:

\begin{figure}[htp]
   \begin{center}
    {\label{}\includegraphics[width=1\textwidth]{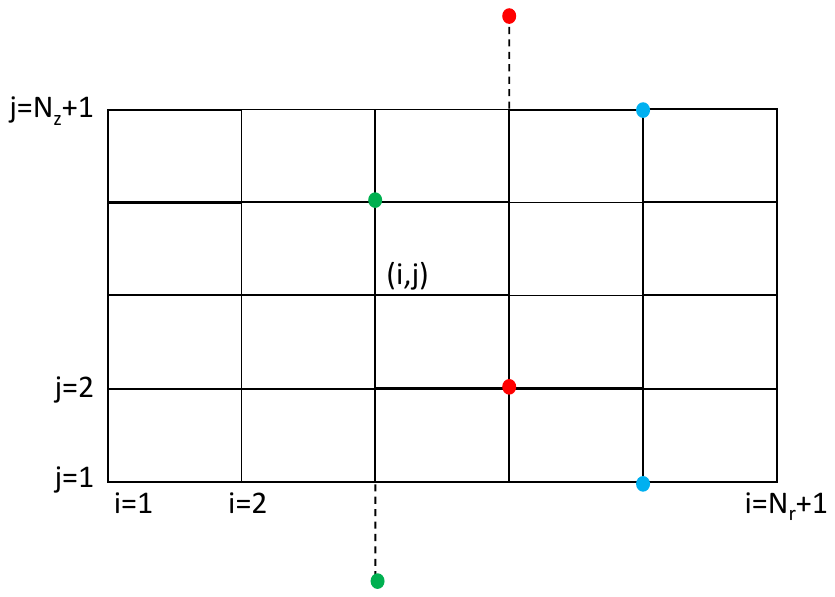}}
  \end{center}
 \caption{Computational grid.}
  \label{fig:grid}
\end{figure}

Let $\Delta\rho$ and $\Delta z$ be the discretisation steps along the $\rho$ and the $z$ axes, respectively, and let $N_r$ and $N_z$ be the number of intervals in the $\rho$ and $z$ direction. Thus, the point $(i,j)$ has coordinates $((i-1)\Delta\rho,(j-1)\Delta z)$, $i=1,\ldots,N_r+1$, $j=1,\ldots,N_z+1$.

We also introduce a global numbering of the nodes in such a way that the node $(i,j)$ is associated to the number $k=(j-1)(N_r+1)+i$.

We now discretise the equation \eqref{eq:abstract} using finite differences. Considering the nature of the boundary conditions, we will write the equation in all the nodes except those on the boundaries $\Gamma_2$ and $\Gamma_4$ where we impose Dirichlet boundary conditions. More precisely, we will consider
\[
  L\theta_{ij} + f(\theta_{ij}) =0 \qquad i=2,\ldots,N_r, \; j=1,\ldots,N_z+1 \, .
\]

\subsection*{Discretisation of the linear term}

To discretise the linear term $L$ we introduce the following second-order approximations of the derivatives
\begin{eqnarray}
\frac{\partial^2\theta}{\partial \rho^2} &\approx& \frac{\theta_{i-1,j} - 2\theta_{ij} + \theta_{i+1,j}}{\Delta \rho^2} \\
\frac{\partial^2\theta}{\partial z^2} &\approx& \frac{\theta_{i,j-1} - 2\theta_{ij} + \theta_{i,j+1}}{\Delta z^2} \\
\frac{\partial\theta}{\partial \rho} &\approx& \frac{\theta_{i+1,j} - \theta_{i-1,j}}{2\Delta \rho}
\end{eqnarray}

Then, with the help of some algebra, the discretisation of \eqref{eq:abstract} becomes
\begin{multline}\label{eq:Lij}
\left( \frac{A}{\Delta\rho^2} - \frac{A}{2(i-1)\Delta\rho^2} \right) \theta_{i-1,j} +
\left( \frac{A}{\Delta\rho^2} + \frac{A}{2(i-1)\Delta\rho^2} \right) \theta_{i+1,j} -
\left( \frac{2A}{\Delta \rho^2} + \frac{2A}{\Delta z^2} \right) \theta_{ij} \\+
\frac{A}{\Delta z^2} \theta_{i,j-1} +
\frac{A}{\Delta z^2} \theta_{i,j+1} + f(\theta_{ij}) = 0 \qquad i=2,\ldots,N_r, \; j=1,\ldots,N_z+1 \, .
\end{multline}

Considering that on $\Gamma_1$ and $\Gamma_3$ we impose periodic boundary conditions, we identify the nodes characterised by indices $(i,1)$ and $(i,N_z+1)$ (shown in blue in the figure above). Moreover, we make the assumption that we can identify the nodes $(i,0)$ with the nodes $(i,N_z)$ (green nodes), and the nodes $(i,N_z+1)$ with the nodes $(i,2)$ (red nodes). Thus, in the matrix $M$ associated to \eqref{eq:Lij} we can identify
\[
  \begin{pmatrix}
  M_{11} & M_{13} & M_{1b} & M_{1I} \\
  M_{31} & M_{33} & M_{3b} & M_{3I} \\
  M_{b1} & M_{b3} & M_{bb} & M_{bI} \\
  M_{I1} & M_{I3} & M_{Ib} & M_{II} \\
  \end{pmatrix}
  \begin{pmatrix}
  \boldsymbol{\theta}_1 \\
  \boldsymbol{\theta}_3 \\
  \boldsymbol{\theta}_b \\
  \boldsymbol{\theta}_I
  \end{pmatrix}
  + f(\boldsymbol{\theta}) = \mathbf{0}
\]
and we can reduce the system to
\begin{equation}\label{eq:system}
  \begin{pmatrix}
  M_{11} + M_{13} & M_{1I} \\
  M_{I1} + M_{I3} & M_{II} \\
  \end{pmatrix}
  \begin{pmatrix}
  \boldsymbol{\theta}_1 \\
  \boldsymbol{\theta}_I
  \end{pmatrix}
  + f\left(\begin{pmatrix}
  \boldsymbol{\theta}_1 \\
  \boldsymbol{\theta}_I
  \end{pmatrix}\right) =
  -
  \begin{pmatrix}
  M_{1b} \\
  M_{Ib} \\
  \end{pmatrix}
  \boldsymbol{\theta}_b
\end{equation}

For simplicity and with obvious choice of notation, we can rewrite \eqref{eq:system} as
\begin{equation}\label{eq:sys2}
\widehat{M} \, \widehat{\boldsymbol{\theta}} + f(\widehat{\boldsymbol{\theta}}) = \mathbf{b} \, .
\end{equation}

\subsection*{Newton's method for the nonlinear system}

To solve the nonlinear system \eqref{eq:sys2}, we consider now the Newton's method: for $k\geq 0$ until convergence, we solve the linear system
\begin{equation}
J_{\widehat{M}}(\widehat{\boldsymbol{\theta}}^{(k)}) \, \boldsymbol{\delta}\widehat{\boldsymbol{\theta}}^{(k)} = - \left(  \widehat{M} \, \widehat{\boldsymbol{\theta}}^{(k)} + f(\widehat{\boldsymbol{\theta}}^{(k)}) - \mathbf{b} \right)
\end{equation}
and set
\begin{equation}
\widehat{\boldsymbol{\theta}}^{(k+1)} = 
\widehat{\boldsymbol{\theta}}^{(k)} + 
\boldsymbol{\delta}\widehat{\boldsymbol{\theta}}^{(k)}
\end{equation}
where $J_{\widehat{M}}(\widehat{\boldsymbol{\theta}}^{(k)})$ is the Jacobian matrix in $\widehat{\boldsymbol{\theta}}^{(k)}$ defined as
\[
  J_{\widehat{M}}(\widehat{\boldsymbol{\theta}}^{(k)})  = \widehat{M} + f'(\widehat{\boldsymbol{\theta}}^{(k)})
\]
and $f'$ is the G\^ateaux derivative of the nonlinear operator $f$ defined as
\[
  f'(\theta_{ij}) = -\left( \frac{A}{((i-1)\Delta r)^2} + K \right) \cos (2\theta_{ij})
  + D \frac{f((j-1)\Delta z)}{(i-1)\Delta r} \sin (2\theta_{ij}) \, .
\]

\end{widetext}


\begin{thebibliography}{50}
\bibitem{explanation}
The wide range of more than 10 orders of magnitude of the possible strength of the polarization reflects the wide range of reported values of coupling constants. We focus on the necessary value of coupling constant that would provide measurable polarization.
\bibitem{remark1}
Due to the physical assumption that the dominant order parameter is the magnetic one, the stiffness energy part of the polarization can be neglected.
\end{thebibliography}
\end{document}